\documentclass[twocolumn,showpacs,preprintnumbers,amsmath,amssymb]{revtex4}
\usepackage[utf8]{inputenc}
\usepackage{graphicx}
\usepackage{subfigure, psfrag, amsmath}
\usepackage{verbatim}
\begin{document}
\preprint{APS/123-QED}

\title{Adaptive self-organization in a realistic neural network model}

\author{Christian Meisel}
\email{meisel@mpipks-dresden.mpg.de}
\author{Thilo Gross}%
\affiliation{%
Max-Planck-Institut f\"{u}r Physik komplexer Systeme, N\"{o}thnitzer Stra\ss e 38, 01187 Dresden, Germany}%

\date{\today}

\begin{abstract}
Information processing in complex systems is often found to be maximally efficient close to critical states associated with phase transitions. 
It is therefore conceivable that also neural information processing operates close to criticality.
This is further supported by the observation of power-law distributions, which are a hallmark of phase transitions. 
An important open question is how neural networks could remain close to a critical point while undergoing continual change in the course of development, adaptation, learning, and more. 
An influential contribution was made by Bornholdt and Rohlf, introducing a generic mechanism of robust self-organized criticality in adaptive networks.  Here we address the question whether this mechanism is relevant for real neural networks. We show in a realistic model that spike-time dependent synaptic plasticity can self-organize neural networks robustly toward criticality.
Our model reproduces several empirical observations and makes testable predictions on the distribution of synaptic strength, relating them to the critical state of the network. 
These results suggest that the interplay between dynamics and topology may be essential for neural information processing.
\end{abstract}


\maketitle

\section{Introduction}
Dynamical criticality has been shown to bring about optimal information transfer and storage capabilities \cite{Beggs2003,Sneppen2005,Legenstein2007} and sensitivity to external stimuli \cite{Haldeman2005,Kinouchi2006} making it an attractive concept for neural dynamics \cite{Bienenstock1998,Freeman2000,Chialvo2006}.
Evidence for dynamical criticality in neural networks is found \emph{in vitro}, in cell cultures and in slices of rat cortex \cite{Beggs2003}, and \emph{in vivo} \cite{Gireesh2008}, where avalanches of neuronal activity observed to follow a power-law size distribution with exponent -1.5. 
On a larger spatial scale of human electroencephalography (EEG), measurements of electrical activity in the brain show that the power spectrum of background activity follows a power-law \cite{Barlow1993,Barrie1996,Freeman2000,Arcangelis2006} which has also been linked to criticality \cite{Novikov1997_2,Arcangelis2006}.
Recently, also measures of phase synchronization were demonstrated to follow power-law scaling in functional magnetic resonance imaging (MRI) and magnetoencephalographic (MEG) data recorded from humans, indicating critical dynamics \cite{Kitzbichler2009}. 

Despite the empirical evidence, few generative mechanisms explaining the purported criticality have been proposed. What is needed is an robust mechanism that drives the system back to the critical state after 
a perturbation. In the theoretical literature, self-organized criticality (SOC), the ability of systems to self-tune their operating parameters to the critical state, has been discussed for a long time \cite{Bak1987}. 
A major new impulse came from the discovery of network-based mechanisms, which were first reported in \cite{Christensen1998} and explained in detail in \cite{Bornholdt2000,Bornholdt2003}. 
These works showed that \emph{adaptive networks}, i.e., networks combining topological evolution of the network topology with dynamics in the network nodes \cite{Gross2008,Gross2009}, can exhibit highly robust SOC based on simple local rules. 

The essential characteristic of adaptive networks is the interplay between dynamics ON the network and dynamics OF the network. In the case of neural networks the dynamics on the networks is the activity of the individual neurons. Dynamics of the network appear in many forms, but include the rewiring of neural connections in the developing brain and synaptic plasticity, the dynamical change of synaptic strength.   
In real world neural networks the formation of connections and all but the fastest mechanism of plasticity are clearly much slower than the dynamics of the neurons. For this reason both types of dynamics can be assumed to take place on separate timescales: On a short timescale the dynamics of neuronal activity occurs in a network with quasi-static topology. Only on a longer timescale, the topology evolves depending on the time-averaged, and therefore also quasi-static patterns of neuronal activity. 

The basic mechanism of adaptive SOC can be explained as follows: Dynamics on networks are in general sensitive to the network topology, the specific pattern of nodes and links. Remarkably, even the dynamics in a single network node may provide information on global topological order parameters, if the node is observed for sufficient time. Thus, the dynamics explores the network making certain global topological properties locally accessible in every network node. In adaptive networks this information can then be utilized by a local topological update rule that slowly drives the network toward criticality. 

The investigation of conceptual models of adaptive SOC \cite{Bornholdt2003,Gross2009} has shown that the presence of this type of self-organization in human neural networks is plausible. 
Independently, robust SOC has recently been demonstrated in neural models \cite{Shin2006,Levina2007,Siri2007}, which also fall into the class of adaptive networks. 
The aim of the present work is to assess whether adaptive SOC can self-organize a realistic model of neural network robustly to criticality. In particular, we consider topological self-organization driven by spike-time dependent synaptic plasticity (STDP). We find that the final state the network approaches is critical, albeit different from the states that were approached in previous models using activity-dependent (homeostatic) mechanism of plasticity.   

\section{Description of the Model}
We consider a network of $N$ leaky integrate-and-fire neurons. In contrast to previous works, which focused on inhibitory interactions \cite{Levina2007}, we study a network of 80\% excitatory to 20\% inhibitory neurons, which is realistic for cortical networks \cite{Braitenberg1991}.
In the absence of a stimulus, the membrane potential, $v_i$, of neuron $i$ follows the equation
\begin{equation}
\label{eqMembrane}
{\rm\frac{d}{dt}}v_i=-\frac{1}{\tau_{\rm m}}[v_i-V_{0}]
\end{equation}
describing an exponential return to the resting potential $V_0$ on a timescale given by $\tau_{\rm m}$.
Whenever a neuron receives an input from a pre-synaptic neuron $j$ we update the membrane potential by adding $[V_{\rm rev}-v_i]g_{\rm c}$ if the pre-synaptic neuron is excitatory, or $-[V_{\rm rev}-v_i]g_{\rm c}$ if the pre-synaptic neuron is inhibitory.    
If the update causes the membrane potential to reach or exceed the threshold $V_{\rm th}$ then the potential is reset to $V_{\rm reset}$ for a refractory period $\tau_{\rm ref}$ after which it evolves again according to Eq.~(\ref{eqMembrane}). 
Upon reaching the threshold the neuron fires a spike, which reaches the connected post-synaptic neurons after a travel-time delay $\tau_{\rm delay}=1.0{\rm ms}$.

Note that spikes can only be fired directly upon receiving an excitatory input. Consequently spikes are fired at times that are integer multiples of the travel-time delay. Thus the travel-time defines a natural timestep which we also use as the timestep of our simulation. Between spikes the membrane potential of a neuron is updated
using the analytical solution of Eq.~(\ref{eqMembrane}).
The system can thus be simulated without suffering from the inaccuracies that are often introduced by numerical integration.  

The topology of the network changes due to spike-time dependent plasticity (STDP) \cite{BiPoo1998,Debanne1998}, 
which is believed to alter the topology of the network on a timescale that is much slower than the spiking dynamics (some hundred milliseconds to some seconds \cite{Markram1997,BiPoo1998}).

Exploiting the timescale separation we proceed as follows: We simulate the dynamics on the network according to the rules described above for a long time $t_{\rm sim}$. Only when the simulation has finished the topology is changed according to an update rule explained below. The timescale $t_{\rm sim}$ is chosen sufficiently long for the system to reach a dynamical attractor. Specifically, we assume that this is the case when the neurons have fired in average $100$ spikes or all activity on the network has stopped. Once the attractor has been reached further simulation does not reveal additional information, as the system remains on the attractor (see e.g.~\cite{Bornholdt2000}). The exact choice of $t_{\rm sim}$ is therefore of no consequence, provided that it is sufficiently long. In the present model, this was confirmed numerically. 

The STDP update rule captures the effect of the temporal order of spikes between pairs of neurons \cite{Markram1997,BiPoo1998,Debanne1998}. Following Refs.~\cite{Pfister2006,Morrison2008}, we model this behavior by introducing internal variables $x_i$ and $n_i$ linked to the activity of a neuron $i$.
The variable $x_i$ encodes the time that has passed since the last spike of neuron $i$, while $n_i$
counts the total number of spikes observed so far.  
At all times $t^{\rm s}$ at which neuron $i$ spikes, both $x_i$ and $n_i$ are increased by $1$.
Between spikes $x_i$ decays with a time constant $\tau_{\rm STDP}$, such that
\begin{equation}
\label{eqTopology}
{\rm\frac{d}{dt}}x_i=-\frac{x_i}{\tau_{\rm STDP}}+\sum_{t^{\rm s}} \delta(t-t^{\rm s}).
\end{equation}
The temporal order of spikes between two neurons $i,j$ can be captured by introducing one more variable $c_{ij}$. When neuron $i$ spikes this variable is decreased by $x_j$, while when $j$ spikes it is increased by $x_i$. Therefore the variable will be increased if the neurons spike in the sequence $i,j$ while it is decreased if the neurons spike in the sequence $j,i$. The increase/decrease is more pronounced in neurons spiking shortly after each other.   

At the end of the simulation run the topology is updated depending on the variables introduced above. 
Depending on the specific question under consideration we use either one of two variants of the topological update rule. The first mimics the formation of neural connections in the developing brain where activity-dependent processes are considered to be a prominent mechanism in shaping topographic maps during development of cerebral connectivity \cite{Innocenti2005,Price2006}. At the time of the topology update, a random $c_{ij}$ is picked. 
If $c_{ij}/(n_i+n_j)$ is greater or equal than the threshold $\theta_{\rm STDP}=0.4$, a new synapse from neuron $i$ to neuron $j$ with $g_{\rm c}=0.15$ is created, if it is smaller than the threshold and a synapse from neuron $i$ to neuron $j$ exists, this synapse is deleted.

For the investigation of the self-organization of synaptic conductance, we use a variant update rule, in which we alter the conductance of a synapse from neuron $i$ to $j$ as $g_{ij}=w_{ij}g_{\rm c}$.
If $c_{ij}/(n_i+n_j)$ is greater or equal than the threshold $\theta_{\rm STDP}$, the weight $w_{ij}$ is increased, if it is smaller, $w_{ij}$ is decreased by a fixed value $\lambda$, unless this would cause $w_{ij}$ to become negative or exceed one.

After the update rule has been applied, we restart the system by assigning random membrane potentials to the neurons and setting two percent of the neurons above threshold. The procedure of simulating the dynamics and then applying the topology update rule is iterated to allow the network topology to relax to a self-organized state.

While our model aims to be as realistic as possible, there are three particulars of real brain networks that we cannot capture: the detailed organization of brain areas, the enormous number of neurons (approx.~$10^{11}$) and the large average number of synapses connecting to every neuron (approx.~$10^4$ for cortical neurons) \cite{Kandel2000}.
While it can be assumed that the detailed organization is only of secondary importance for the questions 
at hand, it is clear that the level of neuronal activity depends on the average number of synaptic inputs.
As we will see in the following, the activity self-organizes to a certain level. Setting the number of synapses to a low/high value therefore causes the synaptic conductances to self-organize to correspondingly high/low levels. Conversely, if we fix the synaptic conductance at a low/high level, the number of synapses self-organizes to a high/low value in turn. Therefore (unrealistically) strong synapses have to be assumed in numerical simulations to keep the number of synapses in a feasible range. The impact of this assumption is discussed below.

Apart from the synaptic conductances $g_{\rm c}=0.15$ all other parameters are set to realistic values. Specifically, $\tau_{\rm m}=30.0{\rm ms}$, $V_{0}=0.0{\rm mV}$, $V_{\rm reset}=13.5{\rm mV}$, $V_{\rm th}=15.0{\rm mV}$, $V_{\rm rev}=33.5{\rm mV}$, $\tau_{\rm delay}=1.0{\rm ms}$, $\tau_{\rm ref}=3.0{\rm ms}$, $\tau_{\rm STDP}=5.0{\rm ms}$  \cite{Diesmann1999,Tsodyks2000,Gewaltig2001}.

\section{Results}
\begin{figure}[!ht]
\begin{center}
\includegraphics[width=0.4\textwidth]{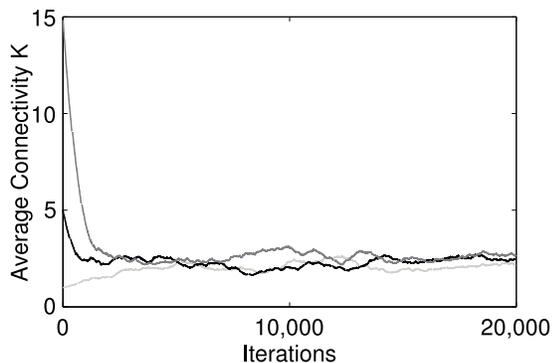}
\end{center}
\caption{Self-organization of network connectivity. Starting from different initial conditions different networks approach the same final level of self-organized connectivity. The exemplary data shown here was obtained in simulations of networks of $N=500$ neurons. While the time-scale separation makes it difficult to relate the iterations directly to biological time the $20000$ iterations in this figure correspond to at least $6000s$.}
\label{figTimeseries}
\end{figure}

As a first test for self-organization, we monitor the connectivity of the model networks, while they evolve under the influence of the STDP update rule. For this purpose the first variant of the update rule is used. Starting with random networks with each neuron having on average synaptic connections to $K$ other neurons, the system approaches a characteristic connectivity $K_{\rm ev}$ independent of initial conditions. A representative set of timeseries is shown in Fig.~\ref{figTimeseries}. Additional investigations (not shown) confirm that $K_{\rm ev}$ is robust against variation of numerical constants such as $t_{\rm sim}$.

\begin{figure}[!ht]
\begin{center}
\includegraphics[width=0.48\textwidth]{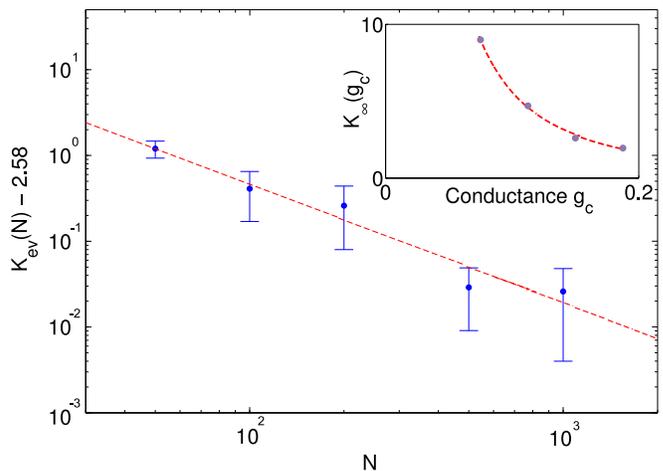}
\end{center}
\caption{(Color online) Scaling of the self-organized connectivity. The final value of connectivity approached in simulation follows a power-law depending on the number of neurons $N$ and synaptic conductances $g_{\rm c}$ (inset).
The values shown were found by averaging over 30000 iterations and 3 estimates of the connectivity $K_\infty$.}
\label{figScaling}
\end{figure}

In order to investigate how our assumptions affect the self-organized level of connectivity, 
we simulate the evolution of networks for different synaptic strength and network sizes.
We find that the value of $K_{\rm ev}$ scales with system size according to the scaling law  
\begin{equation}
K_{\rm ev}-K_{\infty}=aN^{-\beta}
\end{equation}
shown in Fig.~\ref{figScaling}.
The best fit to the numerical observations is provided by the parameter values $K_{\infty}=2.58$, $a=266.1$ and $\beta=1.382$. Computing similar fits for different values of the synaptic conductance we find the scaling law
\begin{equation}
K_{\infty}(g_{\rm c})=bg_{\rm c}^{-\gamma}+c
\label{eqKstar}
\end{equation}
with $b=0.1633$, $\gamma=1.565$ and $c=-0.3916$ (see Fig.~\ref{figScaling} inset). 

In real neural networks, a few tens of simultaneous excitatory post-synaptic potentials are sufficient to elevate the membrane potential from its reset value to spike threshold \cite{Kandel2000}. A typical number is about $20$ inputs corresponding to a conductance of $g_{\rm c}=0.000375$ in our model. Substituting this value into Eq.~(\ref{eqKstar}) we obtain $K_{\infty}\approx 40000$. While this extrapolation certainly provides only a very rough estimate it is reassuring to see that the result is in the order of magnitude of the connectivity observed for cortical neurons \cite{Braitenberg1998}.       

\begin{figure}[!ht]
\begin{center}
\includegraphics[width=0.42\textwidth]{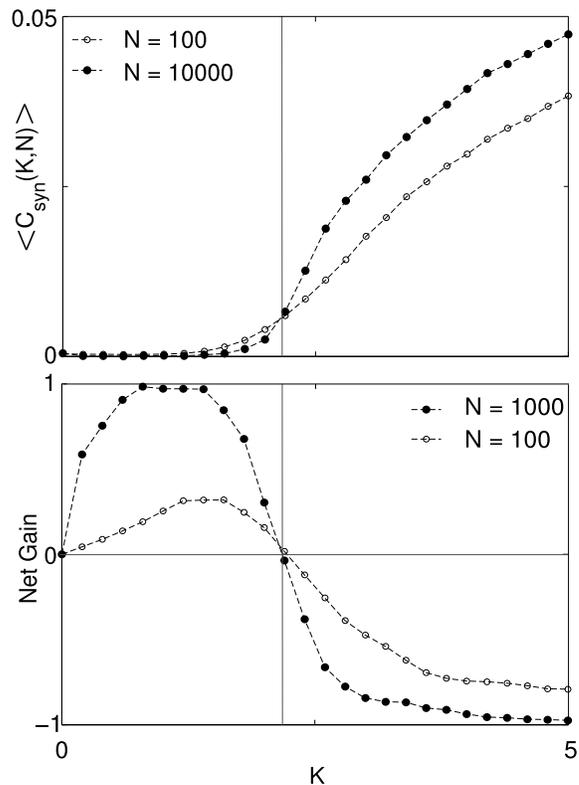}
\end{center}
\caption{Self-organization to the synchronization phase transition. Top: Synchrony of spikes in static networks, measured in terms the order parameter $\langle C_{\rm syn}(K) \rangle$ as a function of the connectivity $K$. The presence of a threshold at $K\approx 2.5$ (line) is indicative of the existence of a phase transition in the limit of large network size. Bottom: Change of connectivity, expressed by the expected average net gain of links per iteration. Links are added if network connectivity is below the threshold and deleted if it is above the threshold, constituting a driving force toward the phase transition.
Samples are averaged over $10^4$ random initial conditions for $N=100$ and $10^3$ random initial conditions for $N=10000$ and $N=1000$.}
\label{figTransition}
\end{figure}

To show that the state approached by the network is critical, we first investigate the dynamics on random networks without synaptic plasticity so that the topology remains static. 
Previous studies \cite{Bornholdt2000,Bornholdt2003} have shown that it is advantageous to quantify the dynamics by defining an order parameter $C_{\rm syn}(K)$ as an average over pairs of neurons of the correlations between neurons $i,j$   
\begin{equation}
C_{i,j}(\tau)=\frac{1}{\tau+1}\sum_{t=t_{0}}^{t_{0}+\tau} \sigma_{i}(t)\sigma_{j}(t),
\end{equation}
where $\sigma_i(t)$ is one if the neuron $i$ spiked at time $t$ and zero otherwise. This quantity is evaluated over a time $\tau$ which we here consider equal $t_{\rm sim}$.
Figure~\ref{figTransition} shows $C_{\rm syn}$ averaged over random network topologies, for different connectivities and network sizes.
This averaged order parameter, $\langle C_{\rm syn} \rangle$, increases at a threshold around $K=2.5$ which becomes more pronounced in larger networks indicating the existence of a continuous phase transition corresponding to the onset of synchrony in neural activity in the limit of large network size. 

As a second step, we compute the net gain of links if the STDP rule were applied. Figure~\ref{figTransition} shows that the STDP increases the connectivity of sparsely connected networks but decreases the connectivity of more densely connected networks. This constitutes an effective force toward the phase transition at $K_{\infty}$.
 
The critical threshold $K_{\rm STDP}$ at which the transition occurs in Fig.~\ref{figTransition} corresponds approximately to the self-organized connectivity $K_{\infty}$ in the dynamic network. An exact match cannot be expected since the evolved networks are no longer random graphs but may exhibit long range correlations in the topology that shift the threshold to higher or lower connectivities.

\begin{figure}[!ht]
\begin{center}
\includegraphics[width=0.48\textwidth]{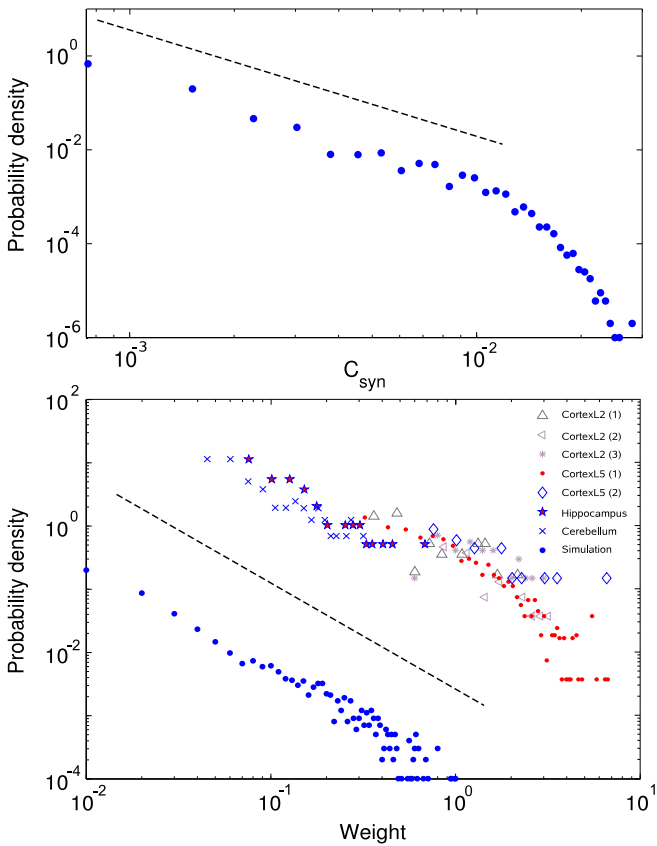}
\end{center}
\caption{(Color online) Power-law distributions in the self-organized state. Top: Distribution of the measure for spike-time synchrony $C_{\rm syn}$ at the self-organized state. The distribution follows a power-law with exponent $-2$ as indicated by the dashed line. Bottom: Distribution of synaptic weights in a self-organized critical state together with probability densities of synaptic weights from experimental data \cite{Barbour2007}. Synaptic weights of experimental data are in ${\rm mV}$, the probability density in ${\rm mV}^{-1}$, respectively. The dashed line indicates a power-law with exponent $-1.5$. 
The data shown were computed in a network with $N=1000$ and $\lambda=0.01$.}
\label{figWeights}
\end{figure}

A hallmark of continuous phase transitions is the power-law distribution of microscopic variables related to the order parameter. We therefore compute the distribution of the spike-time correlation, $C_{\rm syn}$, in the evolving networks. In the self-organized state, we find $C_{\rm syn}$ to be distributed according to a power-law (Fig.~\ref{figWeights}). While our numerical approach imposes strong cut-offs, already the observation of power-law scaling of the variable associated with the order parameter over one decade, provides additional evidence for the dynamical criticality in the evolved state. Our results are reminiscent of recent functional MRI and MEG measurements in human brains, which demonstrated a remarkably robust power-law scaling of two different synchronization measures over a broad range of frequencies and different pairs of anatomical regions \cite{Kitzbichler2009}.

A current question not yet fully resolved by empirical data concerns the distribution of synaptic weights in real neural networks.
To investigate this distribution in our model, we abandon the deletion and creation of synapses and instead switch to the variant STDP update-rule in which the weights of the synapses are increased or decreased. With the variant update rule, we observe that the connectivity, now defined as $K=\sum_{i,j}w_{ij}/N$, approaches the same value $K_{\rm STDP}$ that is found with the boolean update rule.
In the critical state a large fraction of synaptic weights is zero, which is in agreement with empirical evidence \cite{Barbour2007}. In our simulation the exact size of this fraction depends strongly on the number of synapses. 

As shown in Fig.~\ref{figWeights}, the distribution of synaptic weights in the evolved state follows a power-law with exponent $-1.5$. The self-organization to an identical distribution was also observed in further simulations (not shown) using different and/or asymmetric values of $\lambda$ for strengthening or weakening updates, as proposed in \cite{Morrison2008}. Interestingly, a similar scaling behavior regarding the observed power-law for synaptic strengths in our model was found to play a role in models of network pruning \cite{Kurten2008}. The authors relate this observation to the sparseness of networks which is also an outcome of the present model.

For comparison of the self-organized distribution with empirical data, Fig.~\ref{figWeights} also shows measurements of synaptic weights from somatic recordings of different anatomical areas in the brain \cite{Sayer1990,Mason1991,Sjostrom2001,Isope2002,Holmgren2003,Song2005,Frick2008} summarized in \cite{Barbour2007}.
From these recordings the two smallest values are neglected, since measurements of small weights are typically underestimated as they are likely to fall below the detection threshold \cite{Barbour2007}. Comparing the combined data sets to the numerical results reveals a close resemblance. While this could be interpreted as an indication that the brain as a whole is in a self-organized critical state, such a claim is highly questionable, as, at least, the organization into different brain areas is certainly not self-organized. Considered individually, the data-sets curve slightly downwards in the double-logarithmic plot, which is indicative of an exponential distribution \cite{Pajevic2009}. However, statistical tests \cite{Clauset2009} reveal that a power-law relationship can not be excluded at least for CortexL2 (1), CortexL2 (2), CortexL5 (2), Hippocampus and Cerebellum. The exponent providing the best fit for such a power law closely matches the value of $-1.5$ found in our simulations.  
 
\section{Discussion}
In this work we have investigated a realistic model of neural networks, capturing the interplay between the integrate-and-fire dynamics of neurons and spike-time dependent synaptic plasticity. We observed that the networks robustly evolved to a state characterized by the presence of power-laws in the distribution of synaptic conductances and the synchronization order parameter $C_{\rm syn}$.

Our results provide strong evidence for self-organized criticality. In particular they indicate that a previously proposed mechanism, based on the interplay of dynamics and topology in adaptive networks \cite{Bornholdt2000,Bornholdt2003}, can robustly self-organize realistic models of neural networks to critical states. This suggests that this mechanism could also drive the self-organization to criticality recently observed in other models \cite{Shin2006,Levina2007,Levina2009}.

Apart from the higher degree of realism, our model differs from previous works in the nature of the synaptic plasticity. We find that spike-time dependent plasticity drives the network to a critical state that marks the onset of synchronous activity. By contrast, previous models based on activity-dependent update rules, found a 
self-organization toward a different critical state corresponding to an order-disorder phase transition \cite{Bornholdt2000,Levina2007}. Since both types of plasticity have been observed in nature it is promising to study their combined effect in future models. It is interesting to note that the order-disorder transition mainly depends on the average connectivity of the network, while the synchronization transition is much more sensitive to larger topological motifs. The combined effect of activity-dependent and spike-time dependent mechanisms could therefore potentially self-organize the network to both thresholds simultaneously.      

One question that we have not studied in detail in this work concerns the topology that evolves in the model. 
However, the observation of sustained activity in relatively small networks with realistic refractory times suggests the evolution of highly non-random topologies on which activity can propagate similarly to synfire chains.   
Results from an earlier work showed that STDP can reorganize a globally connected neural network into a functional network, which is both small-world and scale-free with a degree distribution following a power-law with an exponent similar to the one for synaptic weights in our model \cite{Shin2006}. 
Similarly, \cite{Siri2007} showed that Hebbian learning can rewire the network to show small-world properties and operate at the edge of chaos. Robust self-organization of scale-free topologies was also observed in a study of spike-time dependent network modifications with node dynamics based on coupled logistic maps \cite{Jost2009}.

Certainly the most important question is if the mechanism studied here is also at work in real neural 
networks. To this end note that our observations were based on the interplay of two well-accepted ingredients: spike-time dependent plasticity and integrate-and-fire neurons. We observed that the coupling of these ingredients yields results that are in good agreement with 
empirical observations. We therefore conclude that also the self-organization to criticality should take place in nature, unless factors exist that specifically disrupt it. The evolution of such factors should be disfavored as critical states are believed to be advantageous for information processing. Furthermore, the basic adaptive mechanism for self-organization, considered here, has by now been observed in several different models. It is therefore unlikely that this mechanism is disrupted by specific details rooted in the biochemistry and biophysics of the neurons. Also, the local nature of the mechanism conveys a high robustness against noise \cite{Bornholdt2000}, which appears in real world networks due to true stochasticity and in the form of external inputs. 
Finally, finite-size effects, which strongly constrain self-organized criticality in many systems, are unlikely to play a role in human cortical networks because of the large number of neurons and synapses. Therefore, the observation of self-organization to criticality in the realistic model, studied here, shows that similar self-organization in real neural networks is likely.

Perhaps the most controversial prediction of the current model is that synaptic weights should follow a power-law. 
Although it is often assumed that the real weight distribution is exponential, to our knowledge, no mechanistic model reproducing the observed distributions has been proposed. Moreover, at least in certain brain areas, the hypothesis that synaptic weights are distributed according to a power-law cannot be rejected on empirical 
grounds. While further investigations are certainly necessary, the mechanism studied here can therefore potentially provide a rationale explaining the observed distributions in these brain areas. 

\section*{Acknowledgments}
CM thanks Jens Timmer and Stefan Rotter for help and discussions.

\bibliography{my}

\end{document}